\documentclass[12pt]{article}

\usepackage{chemformula} 
\usepackage{amsmath}
\usepackage{inputenc}
\usepackage{textcomp}
\usepackage{graphicx}
\usepackage{breqn}
\usepackage{float}
\usepackage[T1]{fontenc} 
\usepackage{times}
\usepackage{authblk}

\newcommand{\Mypm}{\mathbin{\tikz [x=1.4ex,y=1.4ex,line width=.1ex] \draw (0.0,0) -- (1.0,0) (0.5,0.08) -- (0.5,0.92) (0.0,0.5) -- (1.0,0.5);}}%
\title{Differential molecule-cavity mode coupling in  \textit{soft-cavities}}

\author
{Adarsh B Vasista$^{1\ast}$ and William L Barnes$^{1}$}
\affil{$^{1}$Department of Physics and Astronomy, University of Exeter, United Kingdom}
\affil{$^\ast$ \small{E-mail:  a.vasista@exeter.ac.uk} }

\date{}

\begin{document} 


\baselineskip24pt


\maketitle 

\begin{abstract}

The way molecules absorb, transfer, and emit light can be dramatically modified by coupling them to optical cavities. The extent of the modification is often defined by the cavity-molecule coupling strength. Evaluating this coupling strength for different types of modes supported by a cavity is crucial in designing cavities for molecule-cavity coupling. Here we probe a unique multimode cavity, a dielectric microsphere, also called a \textit{soft-cavity}, which supports two distinct types of mode, dark-field scattering (DFS) modes  and whispering gallery modes (WGM). Though seemingly similar, these modes show different characteristics such as spatial electric field profile, resonance line-width etc. We investigated coupling of a mono-layer of J-aggregated dye molecules and a dielectric plastic microsphere using two techniques, far-field excitation and evanescent excitation to generate DFS modes and WGMs respectively. We found that using WGMs we observe a clear signature of strong coupling, whereas with DFS modes we do not. We compared our experimental data to a simple coupled oscillator model and performed finite-element method based numerical simulations to provide a clearer understanding of our experimental findings. 
\end{abstract}

\section{Introduction}
Strong molecule-cavity coupling leads to a radical modification of the energy landscape of the molecule and may result in the emergence of exotic properties such as threshold-less lasing~\cite{5,6}, long-range energy transfer~\cite{7,23}, and tilted ground state reactivity~\cite{8}, among others.
Cavity architectures such as Fabry-Perot cavities~\cite{9,10}, individual nanostructures~\cite{11,12}, nanoparticle arrays~\cite{13,5}, gap-plasmon based cavities~\cite{14}, dielectric nano/microstructures~\cite{1,15} have all been utilized to achieve strong molecule-cavity coupling.  
 
The cavities utilized to achieve strong coupling can be classified into two types: open cavities - where molecules can be adsorbed and desorbed easily, and closed cavities - where there is little room for dynamic molecular movement.
Both open and closed cavities have their advantages and limitations. In the context of applications of strong coupling, open cavities provide the technological advantage of accessibility to the molecular medium.
In this regard, dielectric microspheres, also called \textit{soft-cavities}, are potentialy important.
Such cavities support multiple cavity modes, each with a large figure of merit, and can be easily trapped and moved in a microfluidic environment by optical means \cite{Ashkin1997May}.
Due to an ever-increasing interest in microfluidics with its large  application horizon~\cite{16,17}, there is a clear need to design novel cavities which can be easily integrated with microfluidic systems.
Here we employ two different optical probes to study strong-coupling of molecules with dielectric micro-resonators, with a focus on the different types of information these two probes provide, the probes are: dark-field scattering (DFS), and whispering gallery modes (WGM).

At the individual nano-resonator level DFS has often been used as a tool to investigate the optical modes of the resonator~\cite{Sonnichsen_PRL_2002_88_077402}.
Indeed, DFS has been used to great effect to study strong-coupling, e.g. with the localised plasmon modes of metallic nanoparticles~\cite{Munkhbat_SciAdv_2018_4_eaas9552}.
Whilst the applicability of DFS to strong coupling involving plasmonic nanoparticles is well established, it is not yet clear whether DFS can be applied to dielectric micro-resonators owing to the extent to which the scattering response depends upon the excitation method chosen, typicaly illumination via an optical far-field such as an incident beam of light.
However, dielectric resonators in the several micron size range support spectraly sharp resonances called whispering gallery modes (WGM), and these can be excited by a near-field source, such as that provided by prism-based evanescent excitation. WGMs, due to their strong electric field enhancement and exotic polarization properties~\cite{18}, have applications in e.g. lasing~\cite{21} and enhanced spontaneous emission~\cite{19,20}.
Dark-field scattering by a microsphere also shows spectraly sharp features and these have in the past been studied in the context of Mie scattering~\cite{2,3,4,26,27,28}. Although seemingly similar, as we will show here, the difference between far-field DFS and near-field WGM excitation is critical in understanding their value in probing strong-coupling.
With this motivation in mind we studied strong coupling of a molecular mono-layer of J-aggregated dye molecules with the modes of a dielectric microsphere.
We excited individual microspheres both by evanescent means and through a far-field excitation mechanism, and we found that strong molecule-cavity coupling can be achieved only with WGMs.
We analyzed our experimental data with a simple coupled oscillator model, and also performed finite-element method (FEM) based numerical simulations to help us explain our experimental findings.

\section{Architecture}
\begin{figure}[h!]
\centering
\includegraphics[width=\linewidth]{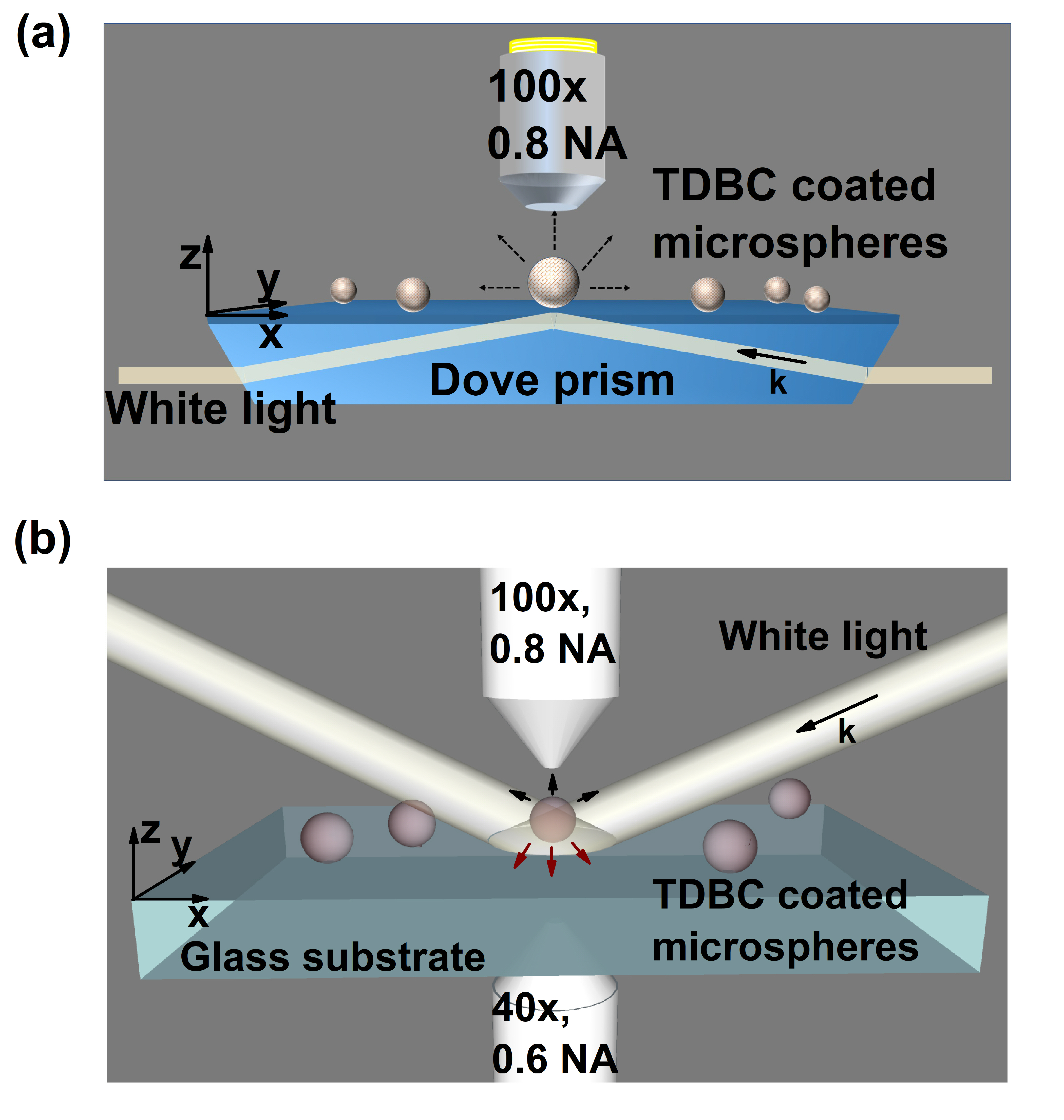}
\caption{\textit{Whispering gallery modes vs. Dark-field scattering}. (a) Schematic representing the configuration to excite whispering gallery resonances of the microsphere through evanescent excitation. Scattered light from individual dye–coated microspheres was collected via the air side through an objective lens. 
(b) Schematic representing the experimental configuration to collect dark-field scattering from an individual mono-molecular layer coated microsphere. The microspheres were excited using white light at an oblique angle and the scattered light was collected using objective lenses. We compared the spectral mode profiles of the scattered light from an individual microsphere both via glass and air side collection. 
}
\end{figure}
To understand the difference between dark-field scattering and whispering gallery modes in the context of their potential as probes of strongly coupled dye molecules  we devised two sets of experiments as depicted in figure 1. \textit{Whispering gallery modes (WGMs)} were excited using evanescent excitation of microspheres through a dove prism as shown in figure 1 (a).
The scattered light was then collected using an air immersion objective lens and analyzed. \textit{Dark-field scattering (DFS)} experiments were carried out by exciting microspheres placed on a glass cover slip with a white-light beam incident at an oblique angle. The scattered light was then collected using an objective lens as shown in figure 1 (b).
We collected the scattered light both on the air-side and through the glass-side so as study any differences in the spectral response.

In both the WGM and DFS configurations, the polarization of the incident light was controlled by placing a linear polarizer in the input path.
Scattering from an individual microsphere was collected by placing a pin-hole in the collection path (see section S1 of the supplementary information for details of the experimental setup).

In both WGM and DFS cases the wavevector of the excitation was kept constant and hence the modes excited inside the microsphere depended primarily on the polarization of the incident light. WGM and DFS modes of the sphere can be classified as transverse electric (TE) and transverse magnetic (TM) modes. TE modes have a tangential electric field profile while the TM modes have a radial field profile. 

\section{Results and discussions}

\subsection{Whispering gallery modes}

\begin{figure}[h]
\centering
\includegraphics[width=\linewidth]{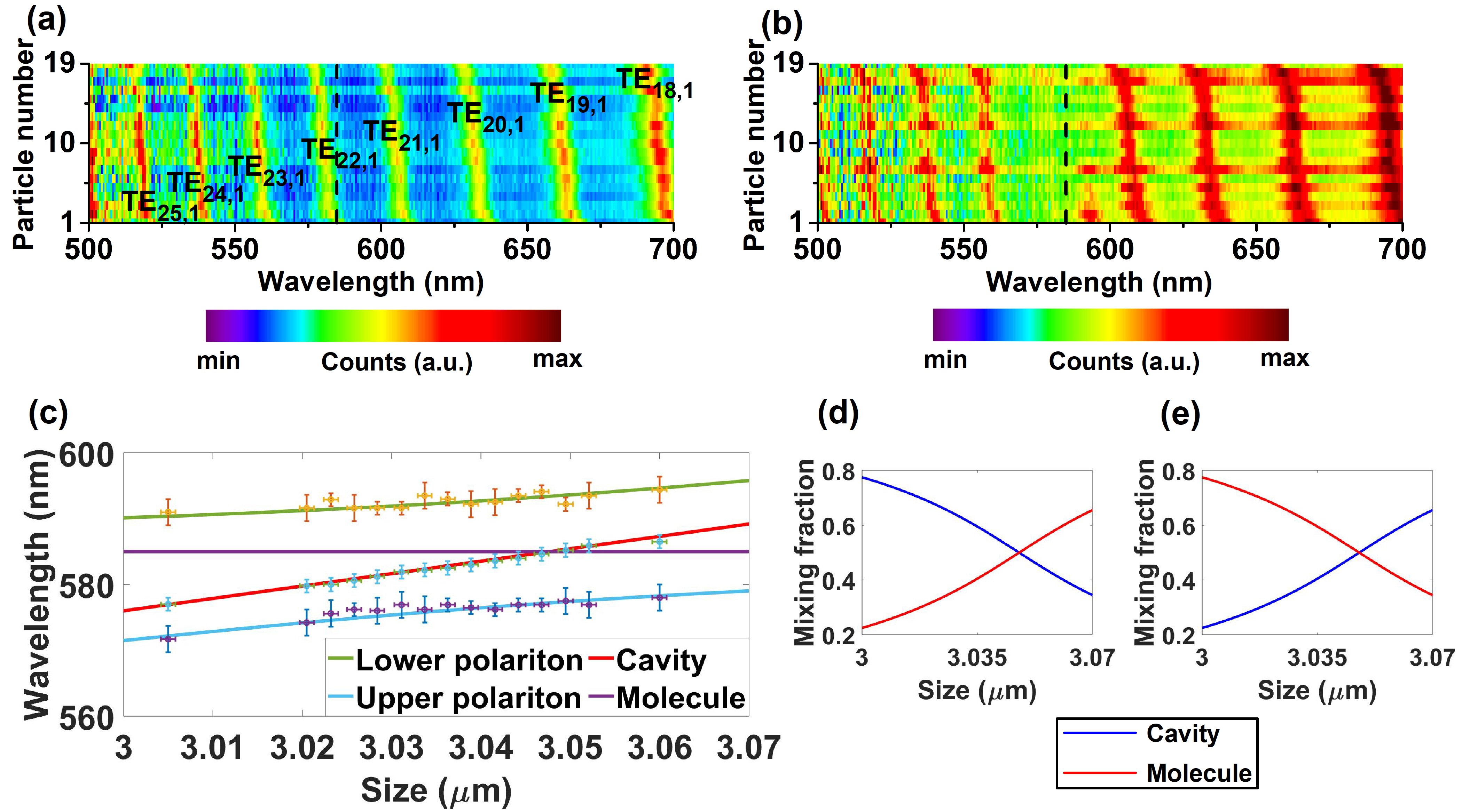}
\caption{\textit{Dispersion of WGMs}. Experimentaly measured dispersion diagram of whispering gallery modes of individual microspheres in the (a) absence and (b) presence of a  molecular mono-layer of PDAC/TDBC. TE modes were excited in the microsphere through evanescent excitation. Mode numbers of WGMs are indicated on the plot. The dashed black line indicates the spectral position of the unperturbed TDBC absorption peak. (c) Calculated dispersion plot using a simple coupled oscillator model to fit the experimental data. (d) and (e) are calculated Hopfield coefficients for the Upper and Lower polaritons respectively. }
\label{fig:false-color}
\end{figure}

To begin with, we excited transverse electric (TE) WGMs of an individual microsphere through a dove prism with s-polarized incident white light.
We performed experiments on a number of microspheres, all nominaly $\sim$ 3 $\mu$m in diameter, but which varied slightly in size, this allowed us to create a dispersion plot as shown in figure 2. The mode numbers were assigned by estimating the size of the microsphere and comparing numerical simulations with the experimental data (see section S2 of supplementary information for details of numerical modeling). Figure 2 (a) shows the dispersion plot of WGMs of microspheres in the absence of a molecular monolayer of PDAC/TDBC (see materials). When a molecular mono-layer of PDAC/TDBC was deposited on the microsphere, as shown in figure 2 (b), the mode TE$_{22,1}$ is seen to overlap with the molecular resonance and split. 

To quantify and understand the strong coupling, the strength of which is indicated by the mode splitting, we modeled the polariton energies using a coupled oscillator model as \cite{7},
\begin{equation}
    \begin{pmatrix}
E_{cavity}-i\frac{\gamma_{cavity}}{2} & -g \\
-g & E_{TDBC}-i\frac{\gamma_{TDBC}}{2}  
\end{pmatrix}
\begin{pmatrix}
\alpha\\
\beta
\end{pmatrix}
= E_{pol}
\begin{pmatrix}
\alpha\\
\beta
\end{pmatrix}
\end{equation}
Where $E_{cavity}$ is the cavity resonance energy, $\gamma_{cavity}$ is the cavity resonance line-width, $E_{TDBC}$ is the molecular absorption energy, and $\gamma_{TDBC}$ is the molecular absorption line-width. The eigenvectors of the coupled oscillator matrix give the Hopfield coefficients ($|\alpha|^2$,$|\beta|^2$) while the eigenvalues ($E_{pol}$) provide the polariton energies \cite{29}. In the present case $\gamma_{TDBC}$=53$\Mypm$1 meV, and $\gamma_{cavity}$= 19$\Mypm$1 meV \cite{1}. Since the molecular absorption energy varies slightly with the refractive index of the substrate \cite{32} we estimated the molecular absorption energy, $E_{TDBC}$, as well as the coupling strength, $g$, by fitting a simple coupled oscillator model to the experimental data. Figure 2 (c) shows the coupled oscillator model fit for the dispersion of the split mode TE$_{22,1}$. The estimated value of the molecular absorption energy was $E_{TDBC}$=2.119$\Mypm$0.004 eV ($\lambda_{TDBC}$=585$\Mypm$1 nm). The coupling strength, $g$, estimated from the coupled oscillator model was 31$\Mypm$2 meV, from which the Rabi splitting $\hbar\Omega_{R}=\sqrt{4g^2-\frac{(\gamma_{TDBC}-\gamma_{cavity})^2}{4}}$ was calculated to be 60$\Mypm$2 meV.
The value of $2g$ (62$\Mypm$4 meV) and the Rabi splitting ($\hbar\Omega_{R}$=60$\Mypm$2 meV) were thus greater than the mean of the line-widths of the cavity (19$\Mypm$1 meV) and molecular absorption (53$\Mypm$1 meV), i.e. more than 36$\Mypm$1 meV, showing the system is in strong-coupling regime \cite{30}. Figure 2 (d) and (e) show the Hopfield coeffecients of the cavity and exciton contributions to the upper and lower polaritons respectively. We compared our experimental results with a simple Lorentzian loss model to ensure that the splitting and the anti-crossing of the WGMs can not be explained by simple multiplication of the absorption of PDAC/TDBC and the WGMs (see section S3 of supplementary information for further details).

To further understand WGM response, we looked at the leakage of WGMs through the glass substrate by illuminating the edge of the glass coverslip so as to excite the waveguide mode of the coverslip.
The waveguide mode in turn excites WGMs of the microsphere drop-cast on the coverslip through evanescent excitation.
We observed strong coupling signatures in the WGM response as collected through glass substrate (see section S4 of supplementary information for further details).

To maximize the molecule-cavity coupling we should ensure minimal polarization mismatch between the modes of the microsphere and the molecular dipoles.
We collected the scattered light from microspheres by exciting both TE and TM modes so as to investigate the orientation of the molecular dipoles.
The TE modes clearly show strong coupling signatures implying a minimal polarization mismatch between the molecular dipoles and the polarization of the mode.
On the other hand the TM modes show no strong coupling signature, implying a polarization mismatch.
(Experimentaly measured dispersion plots of TM WGMs can be seen in section S5 of supplementary information).

\subsection{Dark-field scattering}
\begin{figure}[h]
\centering
\includegraphics[width=\linewidth]{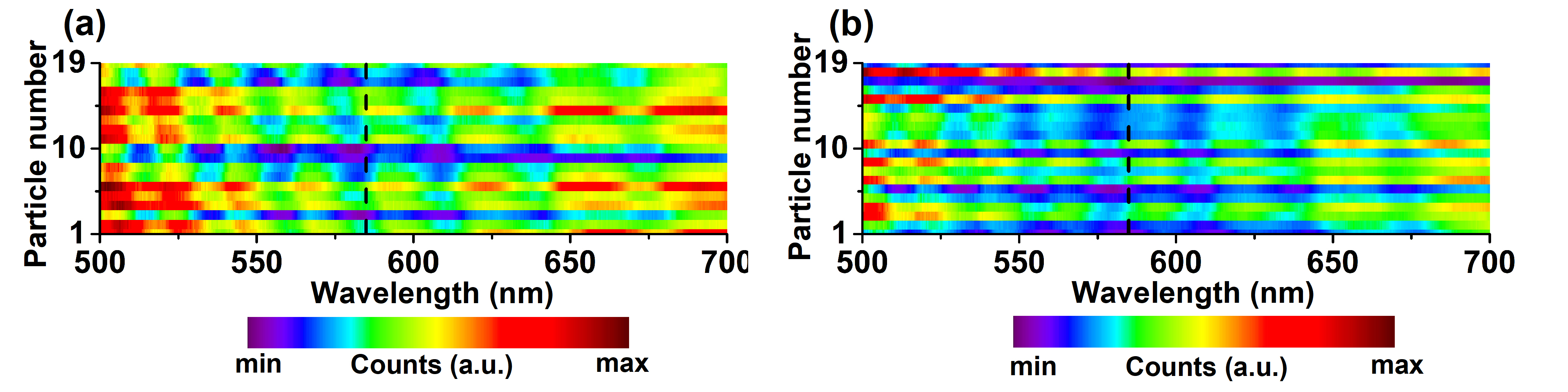}
\caption{\textit{Dispersion of scattering modes, air-side}.  Experimentaly measured dispersion of scattering modes of individual microspheres collected in air-side in the (a) absence and (b) presence of molecular mono-layer of PDAC/TDBC.
TE modes were excited in the microsphere through dark-field excitation at an oblique angle. The dashed black line indicates the position of the unperturbed TDBC absorption peak.
}
\label{fig:false-color}
\end{figure}
With strong coupling of a molecular mono-layer of PDAC/TDBC to a WGM of a microsphere demonstrated, we next focus on the scattering modes of the microsphere.
We probed individual dye-coated microspheres in the dark-field configuration by exciting them at an oblique angle of incidence. We used s-polarized incident light to excite TE modes and the scattered light was collected on both air and glass sides (see section S1 of supplementary information for details of the experimental setup).
Figure 3 (a) shows the dispersion plot of the TE scattered light from individual microspheres collected on air-side in the absence of any dye. There is a clear difference between the spectral positions and linewidths of TE scattering modes and the TE WGMs, the TE scattering modes are red shifted and have a greater linewidth ($\gamma_{scatt}^{air}$=56$\Mypm$3 meV) compared to TE WGMs($\gamma_{WGMs}=$19$\Mypm$1 meV).
Figure 3 (b) shows the dispersion plot of light scattered by microspheres coated with a molecular mono-layer of PDAC/TDBC. There is very little modification by the PDAC/TDBC molecular resonance in the scattering spectra of the microsphere and hence we could not fit a polariton dispersion model to the experimental data. This clearly shows that the system is not in the strong coupling regime. For completeness we also studied the possibility of strong coupling of molecules with TM scattered light and the results can be seen in section S5 of supplementary information. The TM scattered light showed no signature of strong coupling, a similar result to the lack of splitting we see for TM WGMs, as described above.  

To further investigate the possibility of strong coupling with scattering modes of a microsphere, we next collected the scattered light through the glass substrate. Figure 4 (a) shows the dispersion plot of scattering modes of an individual microsphere collected through the glass substrate in the absence of the dye molecules.
\begin{figure}[h]
\centering
\includegraphics[width=\linewidth]{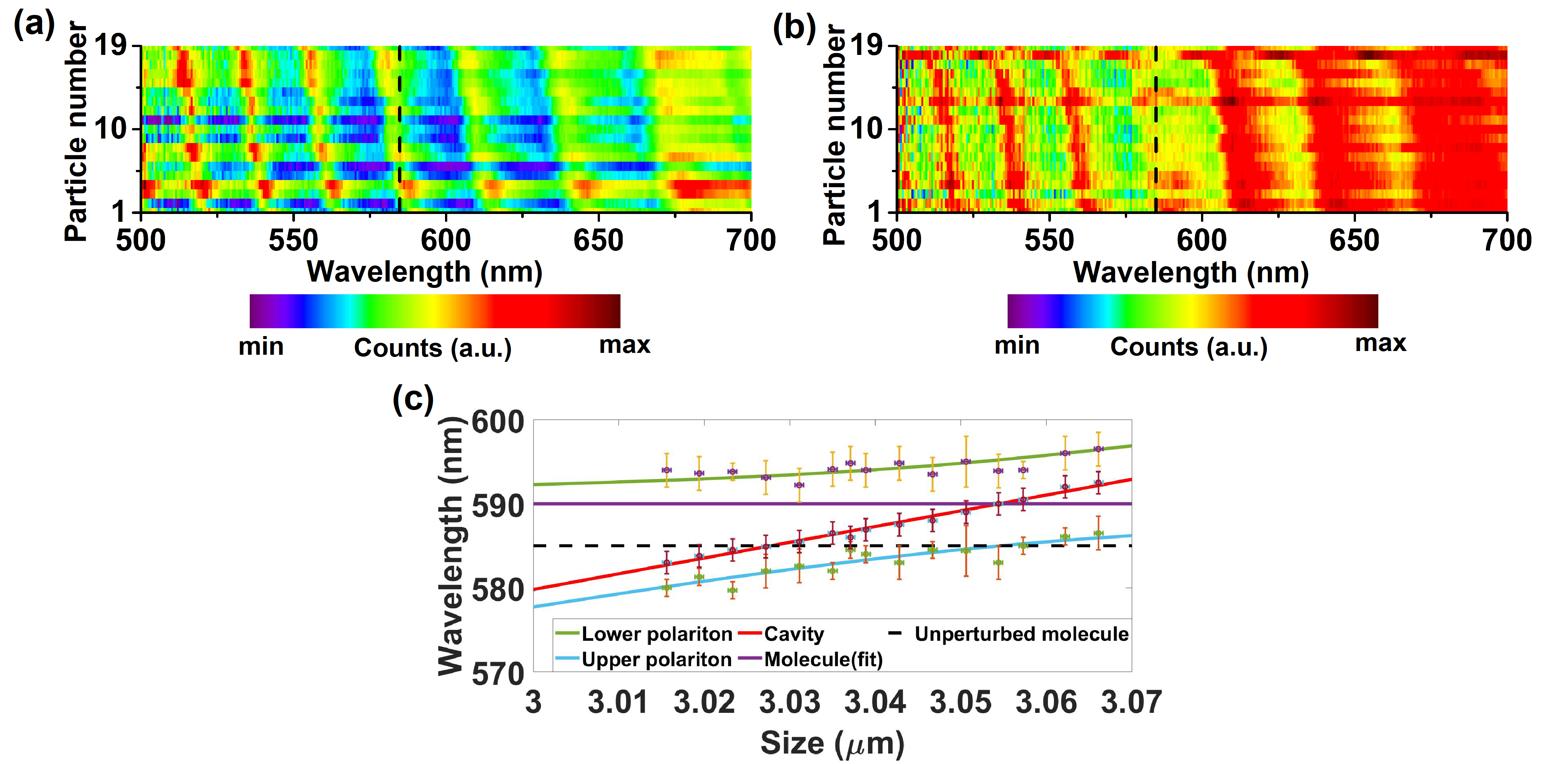}
\caption{\textit{Dispersion of scattering modes; substrate-side collection}. Experimentaly measured dispersion diagram of scattering modes of individual microspheres collected through glass-side in (a) the absence and (b) the presence of molecular mono-layer of PDAC/TDBC. TE modes were excited in the microsphere through dark-field excitation at an oblique angle. The dashed black line represents absorption peak of unperturbed TDBC molecules. (c) Calculated dispersion plot using a simple coupled oscillator model to fit the experimental data. }
\label{fig:false-color}
\end{figure}

Comparing figures 3(a) and 4(a), one can find a clear difference between the spectral profile of scattering modes collected through air and glass sides of the substrate. The scattering modes collected through glass-side are blue shifted and have comparatively narrower line-widths ($\gamma_{scatt}^{glass}$ = 31$\Mypm$3 meV) to the ones collected on the air-side. The scattering modes of a dielectric microsphere are very sensitive to the properties of the substrate and have been widely studied~\cite{2,3,4}. Figures 3(c) and (d) show the scattering modes of an individual microsphere collected through the glass substrate in the absence and presence of a molecular mono-layer of PDAC/TDBC dye respectively. The scattering mode spectraly overlapping with the molecular absorption was broadened and split when a monolayer of dye molecules are coated on a microsphere (see figure 4 (b)), though the extent of splitting is small. 

A closer look at figure 4 (b) reveals that the split modes appear as though the absorption of the TDBC molecules is slightly red shifted. To understand this better, we fit the dispersion of scattering modes with a simple coupled oscillator model (equation 1) to find the absorption maximum and coupling strength, $g$. The simple coupled oscillator model fit for the experimental data is shown in figure 4 (c). The estimated the value of $g$ was 19$\Mypm$2 meV and the molecular absorption was found to peak at 590$\Mypm$1 nm. The value of the coupling strength , $2g$=38$\Mypm$4 meV, was less than the mean of the cavity ($\gamma_{scatt}^{glass}$ = 31$\Mypm$3 meV) and unperturbed molecular absorption ($\gamma_{TDBC}$=53$\Mypm$1 meV) line-widths i.e., less than 42$\Mypm$2 meV. This clearly shows that the system was not in the strong coupling regime. For completeness, the TM scattering modes were also studied and again no signature of strong coupling was observed (see section S5 of supplementary information). This signature of perturbation of scattering modes was obtained in multiple repetitions of the experiments performed to ensure consistency and the dispersion plots are shown in section S6.   

The analysis of dark-field scattering from hybrid J-aggregated molecule-microsphere system raises two important questions:
\begin{itemize}
    \item Why is there a discrepancy between the values of the deduced coupling strengths for WGMs and scattering modes? 
    \item Why does the absorption of the J-aggregate appear to be red shifted in the scattering data?
    \end{itemize}

\subsection{Simulating strong coupling in microspheres}

\begin{figure}[h]
\centering
\includegraphics[width=\linewidth]{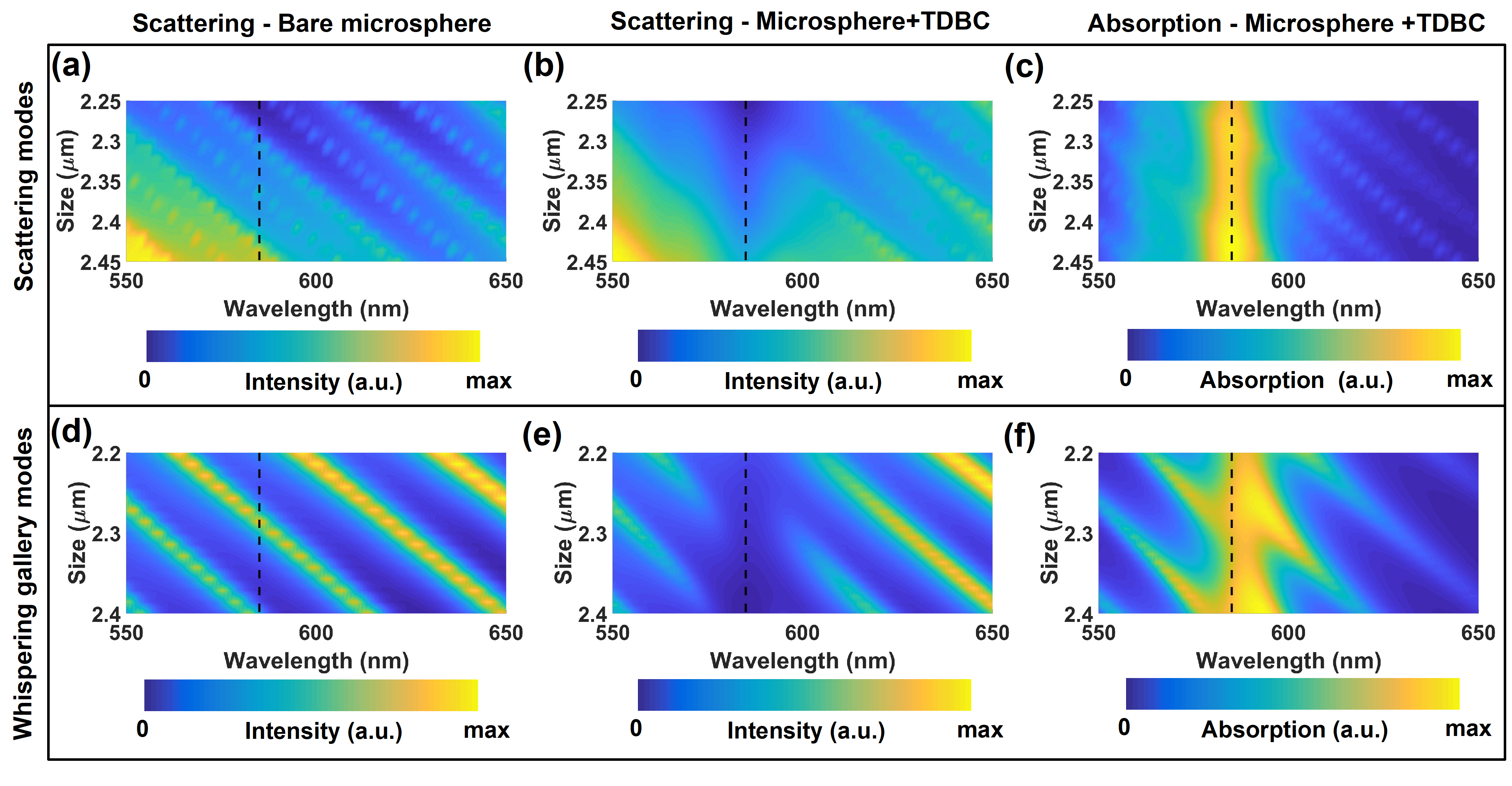}
\caption{\textit{Scattering modes v/s Whispering gallery modes}.(a) and (b) are the calculated dispersion of far-field scattering modes of an individual microsphere of size $\sim$ 2.3 $\mu$m in the absence and presence of a molecular mono-layer of PDAC/TDBC respectively. (c) Calculated dispersion of absorption of the molecule-microsphere hybrid system excited through far-field excitation. (d) and (e) are the calculated dispersion of TE whispering gallery modes of an individual microsphere of size $\sim$ 2.3 $\mu$m in the absence and presence of a molecular mono-layer of PDAC/TDBC respectively. (c) Calculated dispersion of absorption of the molecule-microsphere hybrid system excited through evanescent excitation. Dashed black line represents the absorption maximum of unperturbed PDAC/TDBC molecules.}
\label{fig:false-color}
\end{figure}

To further our understanding of molecule-cavity coupling in microspheres we performed numerical simulations by exciting microspheres through both evanescent and far-field means. First, we calculated far-field scattering from an isolated microsphere in vacuum using 3D numerical simulations. Due to computational limitations: (i) we consider microspheres of smaller size ($\sim$ 1.6 $\mu$m and $\sim$ 2.3 $\mu$m) than the ones used in experiments; (ii) we ignore the effect of the substrate. The simulations performed subject to these approximations provide a proof-of-principle of a general trend in molecule-cavity coupling strength with scattering modes and the results can then be extrapolated to $\sim$ 3$\mu$m microsphere on glass substrate using 2D numerical simulations (see section S2 for details on modeling). Figure 5 (a) shows the dispersion of scattering spectra of a bare microsphere. Since the system is sphericaly symmetric, we excite both TE and TM modes. Figure 5 (b) shows the dispersion of scattering spectra after a molecular mono-layer of PDAC/TDBC was deposited on the microsphere. The scattering spectra are smudged, specificaly at the higher energy side, and show a dip near the absorption maximum of the unperturbed PDAC/TDBC. However, it is difficult to conclude positively on strong coupling by observing the scattering spectra as these features could also arise from the incoherent interaction between molecular dipoles and the scattering modes (convolution of molecular absorption and scattering spectra). To better understand the hybrid system, we calculated the dispersion of the absorption spectra of the hybrid system, shown in figure 5 (c). One can see that the system is in the weak coupling regime as the absorption spectra of the hybrid system are a simple convolution of molecular absorption and the scattering modes. To draw a comparison between soft-cavities and Mie exciton-polaritons in high-index nanoparticles, we calculated the dispersion of scattering modes in a silicon nanoparticle of size $\sim$ 80 nm in the presence and absence of molecular mono-layer of PDAC/TDBC and the results are shown in section S7 of supplementary information.

Next we calculated WGMs for a similar size range of microspheres using 2D numerical simulations. Figure 5 (d) and (e) show the dispersion of TE WGMs excited in the microsphere in the absence and presence of a molecular monolayer of PDAC/TDBC respectively. We can clearly see a splitting and an anti-crossing of the WGMs spectraly resonant with the molecular absorption. To further clarify the strong coupling aspect, we calculated the dispersion of the absorption spectra of the hybrid system, shown in figure 5 (f). The dispersion shows a splitting and anticrossing of modes around 585 nm indicating the system is indeed in the strong coupling regime. By comparing figures 5 (a)-(c) and 5 (d)-(f) one can confirm that we can achieve molecular mono-layer strong coupling with WGMs but can not do so with the scattering modes, this is consistent with our experimental observations. We have also repeated these calculations for a microsphere of size $\sim$1.6 $\mu$m and the results are provided in section S7 of the supplementary information, they confirm this conclusion.

\begin{figure}[h]
\centering
\includegraphics[width=\linewidth]{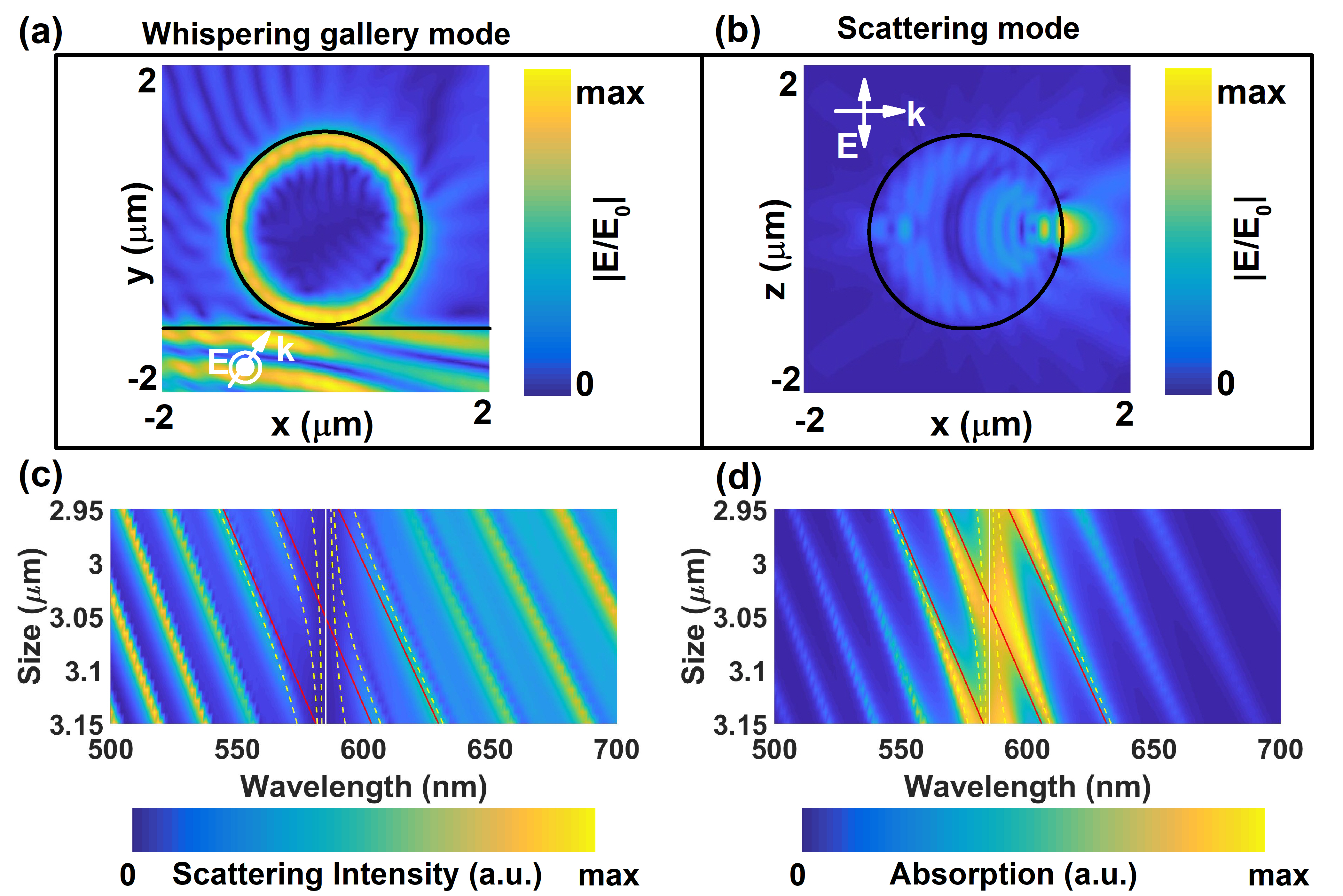}
\caption{\textit{Simulating molecule-cavity coupling in soft-cavities}. Numericaly simulated electric field profiles of an individual microsphere of size 2.36$\mu$m excited through (a)  evanescent means and (b) far-field means respectively. (c) and (d) are the dispersion of scattering and absorption of TE WGMs of an individual microsphere of size $\sim$ 3 $\mu$m coated with a molecular monolayer of TDBC respectively. The calculated polariton dispersions using coupled oscillator model is superimposed. The yellow dashed line represents the polaritons, red line represents the cavity mode, and white line represents molecular absorption of unperturbed PDAC/TDBC.}
\label{fig:false-color}
\end{figure}

To understand the difference between scattering modes and the WGMs in the context of strong coupling we next look at the electric field profiles of the modes involved. Figure 6 (a) shows the simulated electric field profile of a WGM (TE$_{16,1}$) with a sphere of size 2.36 $\mu$m excited using evanescent means at a wavelength of 603 nm. The profile clearly shows the electric field maximum along the periphery of the microsphere. This facilitates maximum interaction between the molecular dipoles and the cavity mode, resulting in strong coupling. On the other hand, figure 6 (b) shows the simulated scattered electric field profile of a scattering mode of the microsphere of size 2.36 $\mu$m (illumination wavelength = 582 nm, a scattering maximum). The microsphere, here, acts as a microlens to the incoming beam and thus focuses the beam producing a photonic nanojet \cite{34,35}. The number of molecules coupled to the mode will be extremely small in this configuration and hence the value of molecule-field coupling strength will be low. It is due to this mismatch that no strong coupling signature was observed with the scattering modes. Similar electric field profiles were obtained for a microsphere of size $\sim$ 3 $\mu$m excited through evanescent and far-field means (see section S7 of supplementary information). 

After distinguishing scattering and whispering gallery modes with respect to molecule-cavity coupling, we next look at the modification of the scattering and absorption properties of the molecule-WGM hybrid system. Figure 6 (c) shows the calculated dispersion plot of WGMs of an individual microsphere of size $\sim$ 3 $\mu$m coated with a molecular mono-layer of PDAC/TDBC (see section S2 of supplementary information for details on numerical modeling). We have included a wider range of microsphere size here to gain a better understanding of the dispersion behaviour of the WGMs. Due to the presence of a molecular mono-layer, modes TE$_{23,1}$, TE$_{22,1}$, and TE$_{21,1}$ were perturbed. In the range of microsphere sizes plotted one can clearly see that the mode TE$_{22,1}$ splits into two polariton branches and undergoes anti-crossing. To model mutli-mode coupling to a single excitonic resonance, we used a 6x6 coupling matrix with 3 fold degenerate exciton energy\cite{33,36}. The coupled oscillator equation reads
\begin{equation}
   \bigoplus_{j=1}^{3} \begin{pmatrix}
E_{j}-i\frac{\gamma_{j}}{2} & -g_{j} \\
-g_{j} & E_{TDBC}-i\frac{\gamma_{TDBC}}{2}  
\end{pmatrix} 
    \begin{pmatrix}
    \alpha_{1} \\  \alpha_{2} \\  \alpha_{3} \\  \alpha_{4} \\  \alpha_{5}\\  \alpha_{6}
    \end{pmatrix} = E_{pol}
    \begin{pmatrix}
    \alpha_{1} \\  \alpha_{2} \\  \alpha_{3} \\  \alpha_{4} \\  \alpha_{5}\\  \alpha_{6}
    \end{pmatrix}
\end{equation}
where $E_{j}$ are the cavity resonance energies, $\gamma_{j}$ are the cavity linewidths, $g_{j}$ are the coupling strengths, $\alpha_{1}-\alpha_{6}$ are eigen vectors, and $E_{pol}$ are the polariton energies. The value of $g_{j}$ were adjusted to fit the numericaly simulated dispersion of WGMs (see section S8 of supplementary information for further details on modeling). The calculated polariton dispersion is superimposed on the numericaly simulated scattering of WGMs in figure 6 (c). Strong molecule-cavity coupling not only alters the scattering behaviour of the hybrid system but also its absorption. The simulated dispersion of molecular absorption of the hybrid system is shown in figure 6 (d). Since the dielectric microsphere shows minimal absorption, figure 6 (d) represents the molecular absorption of PDAC/TDBC when coupled to the WGMs of a microsphere. We can notice the splitting and anti-crossing of the mode TE$_{22,1}$ while modes TE$_{23,1}$ and TE$_{21,1}$ are also affected. Due to this multi-mode coupling, the absorption landscape of the molecule has been altered. The coupled oscillator matrix of equation 2 was used to fit the numericaly simulated absorption modes of the hybrid system and is shown in figure 6 (d). The dark-field scattering modes probed this modified absorption due to vacuum strong coupling of PDAC/TDBC with WGMs. In the size range of 3$\mu$m-3.07$\mu$m (the experimentaly probed size range) the polaritons, specially the lower polariton, are almost dispersionless (see section S8 of supplementary information for further details). In addition to which, the lower polariton branch is more intense than the upper polariton branch. This feature itself manifests as an equivalent redshift in the absorption when probed with the dark-field scattering modes. 



\section{Conclusions}
In summary, we have studied strong molecule-cavity coupling through both dark-field scattering and the excitation of whispering gallery modes of an individual microsphere. We show that WGMs are excellent candidates to achieve strong coupling, even when only a molecular mono-layer is used; in contrast, the use of dark-field scattering does not yield any signature of strong coupling. Our experimental data are supported by numerical simulations. The results presented here open a new avenue in understanding strong molecule-cavity coupling in dielectric microstructures, and will find relevance in designing cavities for molecular strong coupling, especialy for fluidic systems.

\section*{Materials}
Polystyrene microspheres of size $\sim$ 3 $\mu$m were coated with a molecular mono-layer of J-aggregated 5,5',6,6'-tetrachloro-1,1'-diethyl-3,3'-di(4–sulfobutyl)-benzimidazolocarbocyanine (TDBC) dye using layer-by-layer (LBL) deposition method~\cite{Bradley_AdvMat_2005_17_1881}. In a typical LBL deposition scheme, a layer of oppositely charged polyelectrolyte will be coated on the surface of a microsphere to bind the dye molecule.
Here we used a combination of positively charged poly(diallyldimethylammonium chloride) (PDAC) and negatively charged poly(sodium 4-styrenesulfonate) (PSS) as polyelectrolyte binders on an aliphatic amine functionalized microsphere.
Dye-coated microspheres were dropcast on a glass coverslip and studied using a custom built microscope. 

\section*{Acknowledgements}
We thank Wai Jue Tan for his help in preparing samples, and Philip Thomas for fruitful discussions on various strong coupling criteria. The authors acknowledge funding from the European Research Council through the Photmat project (ERC-2016-AdG-742222 www.photmat.eu).Data in support of our findings are available at:

https://ore.exeter.ac.uk/repository/handle/XXX

\section*{Supplementary Information}

A document containing information about simulation strategy, TM polarized dark field scattering spectra from microspheres, electric field profiles and calculated dispersion plots of microspheres are available free of charge. Research data are available from the University of Exeter repository at https://doi.org/xxxx

\section*{Author contributions}
ABV conceived the idea, planned the project, and performed all the experiments and simulations. ABV and WLB analyzed the data. ABV wrote the paper with inputs from WLB.


\bibliography{reference_pra}
\bibliographystyle{unsrt}
\end{document}